# Schottky Barrier MOSFET Enabled Ultra-Low Power Real-Time Neuron for Neuromorphic Computing

Shubham Patil, Jayatika Sakhuja, Ajay Kumar Singh, Anmol Biswas, Vivek Saraswat,
Sandeep Kumar, Sandip Lashkare, Udayan Ganguly
Electrical Engineering, IIT Bombay, Bombay, India. Email: udayan@ee.iitb.ac.in

**Abstract**
Energy-efficient real-time synapses and neurons are essential to enable large-scale neuromorphic computing. In this paper, we propose and demonstrate the Schottky-Barrier MOSFET-based ultra-low power voltage-controlled current source to enable real-time neurons for neuromorphic computing. Schottky-Barrier MOSFET is fabricated on a Silicon-on-insulator platform with polycrystalline Silicon as the channel and Nickel/Platinum as the source/drain. The Poly-Si and Nickel make the back-to-back Schottky junction enabling ultra-low ON current required for energy-efficient neurons.

## Introduction

In the internet of things (IoT) era, data-centric computation is gaining significant interest due to its high energy efficiency. The traditional computers are based on von Neumann architecture, where data and processing units are located separately. Hence, there is significant energy consumption during data transfer between memory and processing unit [1].

Among various alternate technologies, neuromorphic computing is gaining significant interest as its architecture is inspired from the human brain, where memory and processing units are not separate. A spiking neural network (SNN) is a 3rd generation neural network that offers energy-efficient computation by realizing brain functionalities [2]. Synapses and neurons are the basic building blocks of the SNN [3]. Various silicon and non-silicon-based synapse and neuron demonstrations exist in the literature [4-6]. Though the main functionality of synapse and neuron is demonstrated, a complete circuit realization requires the current-to-voltage (V(I)) converter in the synapse implementation. The I-V converter ensures that the synapse type and array are neuron-design-independent by avoiding the loading effect. Similarly, Voltage controlled current source (I(V)) is required in the neuron implementation, which makes the neuron independent of synapse type and array. Hence, the I(V) and V(I) are essential to make both the synapse and neuron modular (Fig. 1).

The Voltage controlled current source (I(V)) is generally implemented by applying the voltage to a typical metal oxide semiconductor field effect transistor (MOSFET). However, the ON currents (mA) of the MOSFET are very large, making the neuron very fast (MHz) (as compared to <kHz for biological neurons) [5]. To lower the power consumption, the sub-threshold region of MOSFET is utilized with low ON current [7]. However, this makes the V-I conversion non-linear making adding additional circuit requirements on neurons (Table 1). To resolve this challenge, in this paper, we propose and demonstrate a Schottky barrier (SB) MOSFET for a voltage-dependent ultra-low current source (I(V)) for power-efficient real-time neurons.

In this paper, first, we demonstrate the SB-MOSFET on the SOI platform with Gd2O3 as box oxide. The polycrystalline Silicon is used as channel material, and Nickel (Ni)/Platinum (Pt) is used as a source/drain to make the Schottky barrier. Next, we demonstrate the SB limited ultra-low current controlled by the top and bottom gate voltage. Finally, a real-time neuron (<kHz) is demonstrated using the ultra-low current output from the SB MOSFET, and a Spiking Neural Network (SNN) is demonstrated to show software equivalent learning accuracy. Such a voltage to the ultra-low current device is essential to implement a real-time neural network that can effectively mimic human brain functionality to solve intricate real-time applications.

## Device Fabrication and characterization

The fabricated SB MOSFET schematic and the detailed process flow are shown in Fig. 2. First, the epitaxial $Gd_2O_3$ layer is deposited on Si (111) with resistivity 0.001 Ω-cm in the RF magnetron sputter of AJA International using a $Gd_2O_3$ target with 99.999% purity [8]. Second, the active area is defined using the deposition and lift-off of an undoped Si layer (~40 nm) using RF sputtering. Next, the sample is exposed to rapid thermal annealing (RTA) at 400 °C/1 min in $N_2$ ambient to improve the mobility of the channel layer. Third, Source and drain are defined using Ni/Pt stack deposition and lift-off using AJA 6-target e-beam evaporator. Forth, the high-k Hafnium oxide ($HfO_2$, 16 nm) as gate insulator is deposited using atomic layer deposition (ALD), followed by gate metal (Tungsten (W)) deposition using DC sputtering. Finally, the S/D contact pads are defined by wet etching of $HfO_2$ using BHF.

The fabricated structure is structurally and electrically characterized. The surface HRSEM image of the fabricated SB MOSFET is shown in Fig. 3. The devices with gate lengths ($L_G$) of 10, 20, 25, and 75 μm with fixed Width (W) = 33 μm are used in this work. The capacitor with a value of 10 pF and 4.7 nF is used for neuron characterization.

## Results and discussions

*A. Electrical characterization*

*I. Transfer and output characteristics*: Fig. 4 shows the transfer (Fig. 4a) and output characteristics (Fig. 4b) of the SB MOSFET for $L_G/W = 20/33$ μm using the Si (p++) substrate as the back gate (BG) with $Gd_2O_3$ oxide. We observe the good back gate control with low leakage current (pA) through the oxide (Fig. 4c), confirming its good quality. Fig. 5 shows the transfer (Fig. 5a) and output characteristics (Fig. 5b) of the SB MOSFET for $L_G/W = 10/33$ μm with top gate (TG) control with $HfO_2$ as a gate oxide. We have shown typical control of a MOSFET with current limited by SB with $V_{GS}$ and $V_{DS}$ dependence. Next, the extracted device performance parameters, (a) $g_m$ (Fig. 5c), (b) $\mu_{eff}$ (Fig. 5d), and (c) $R_{SD}$ (Fig. 5e), are shown. The low current of fabricated MOSFET is due to (1) the Schottky barrier at the S/D end and (2) the high resistivity of the Silicon channel layer.

Next, we have shown the drain current tunability using the back gate bias ($V_{BG}$) for a fixed tog gate bias ($V_{TG}$), as shown in Fig. 6a. This shows the threshold voltage tuning of the MOSFET with BG bias. Furthermore, Fig. 6b shows the impact of $L_G$ scaling on device performance. Drain current is inversely proportional with $L_G$ as typical to traditional bulk MOSFET (Fig. 6c). This fabricated "voltage-dependent ultra-low current source" can enable compact and energy-efficient Neuromorphic architectures. Next, we show neuronal behavior.

*II. Neuron demonstration:* The circuit used for the neuron demonstration is shown in Fig. 7a. The voltage measurement across the capacitor directly using oscilloscopes is challenging, as it has a low maximum impedance (1 MΩ). To circumvent this problem, the measurement is carried out using a buffer (input resistance > 10 TΩ) between (C) of 4.7 nF is used for this measurement. Fig. 7b shows the output voltage ($V_{out}$) versus time for different $V_{TG}$s at fixed $V_D = 1.5$V. The capacitor charging time decreases with an increase in $V_{TG}$. The extracted frequency versus $V_{TG}$ is shown in Fig. 7c for (1) $V_D$ =1.5 V and 2.5 V at fixed C = 4.7nF and (2) for two different C values of 10 pF and 4.7 nF at $V_D = 2.5$ V. The threshold voltage ($V_{th}$) is set at 0.6 V. we observe the frequency tunability with a change in $V_D$. Further, ~5.4x frequency improvement is shown for the 10 pF case. Hence, we demonstrate the real-time neuron (<kHz) akin to biological neuron frequency using the ultra-low current output from the SB MOSFET.

Next, we implemented a two-layer feedforward (16 ×3) SNN using a supervised learning algorithm with Leaky Integrate and Fire (LIF) neurons and plastic synapses in MATLAB to solve Fisher's Iris classification problem [3]. The gate voltage ($V_{TG}$)-dependent SB current is modeled by a voltage-dependent current source ($I_{in} = f(V_{TG})$) (Fig. 7d). A software equivalent accuracy of 96.7% is shown as compared to the ideal LIF neuron for the Fisher Iris classification data (Fig. 7e).

## Conclusion

We demonstrated the SB-MOSFET with ultra-low current controlled by the top and bottom gate voltage on the SOI platform with $Gd_2O_3$ as box oxide. The ultra-low current output from the SB MOSFET was used to demonstrate the real-time neuron (<kHz) akin to biological neuron frequency. Low controllable current sources are critical to reduce the capacitive area for implementing the sub-kHz time scale neurons for processing real-world signals.

## Acknowledgment

This work is partially supported by DST and MeitY.

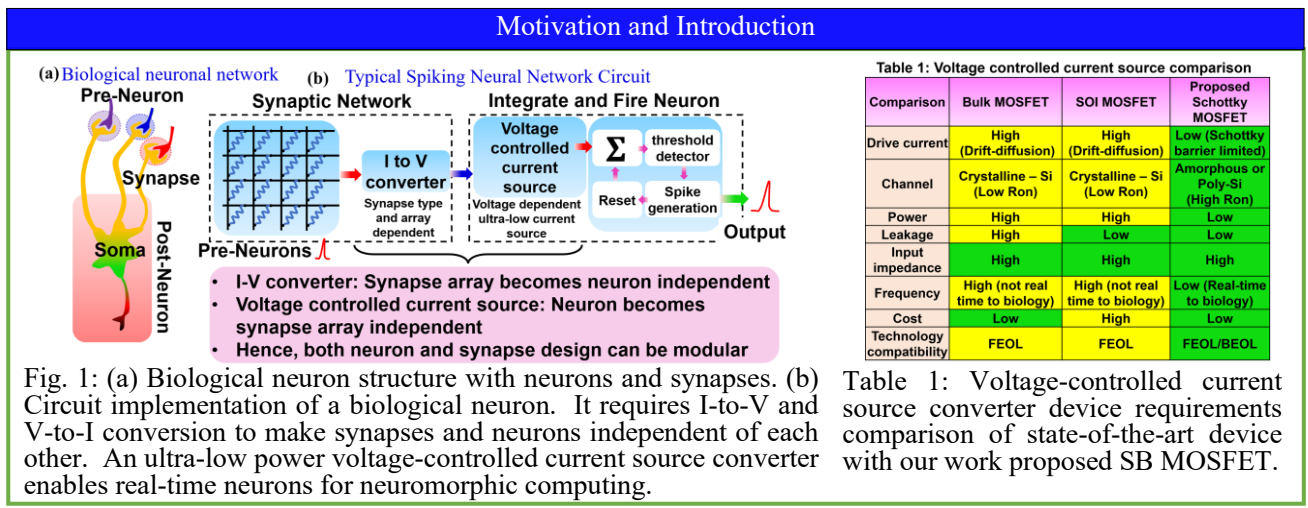

Fig. 1: (a) Biological neuron structure with neurons and synapses. (b) Circuit implementation of a biological neuron. It requires I-to-V and V-to-I conversion to make synapses and neurons independent of each other. An ultra-low power voltage-controlled current source converter enables real-time neurons for neuromorphic computing.

Table 1: Voltage-controlled current source converter device requirements comparison of state-of-the-art device with our work proposed SB MOSFET.

## Fabrication and physical characterization

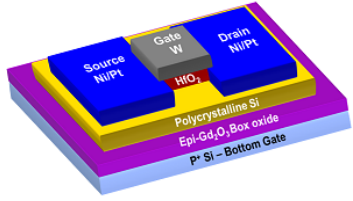
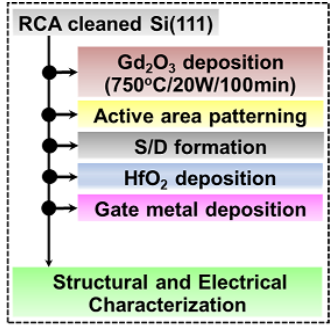
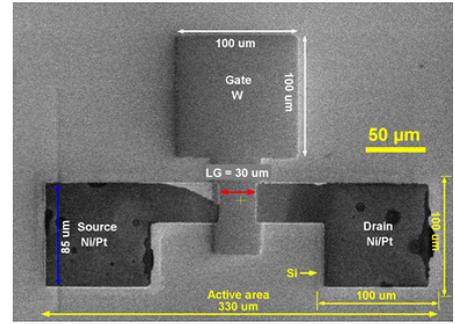

Fig. 2: Schematic and process flow of the fabricated SB MOSFET used in this work. Ni/Pt stack is used as S/D contact.

Fig. 3. Surface scanning electron microscope (SEM) image of the fabricated SB MOSFET.

## Electrical characterization: Ultra Low Current SB MOSFET

### Back Gate (BG) control
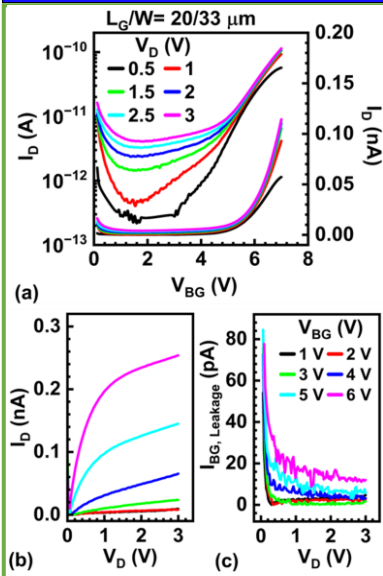

### Top Gate (TG) control
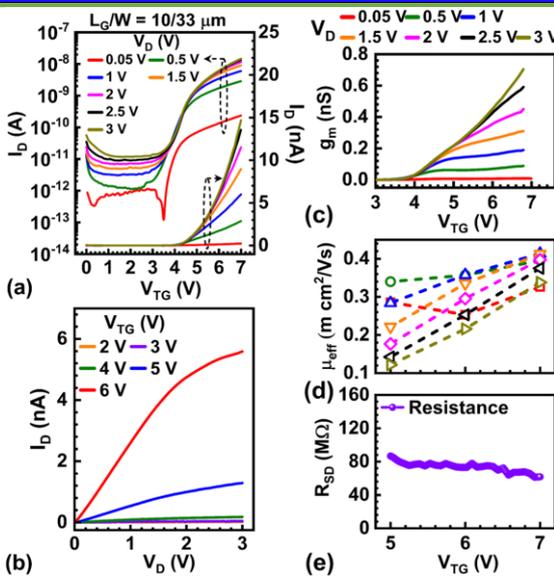

### Double gate & Scaling
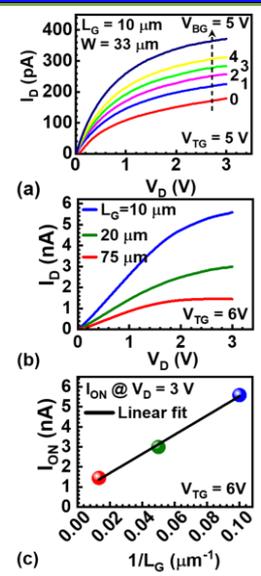

Fig. 4 (a) $I_D$-$V_{BG}$, (b) $I_D$-$V_D$, and (c) $I_{BG,Leakage}$-$V_D$ of the MOSFET with back gate (BG). Good gate control with low box oxide leakage current (pA) confirming its good quality.

Fig. 5 Measured transfer (a) and output (b) characteristics of the SB MOSFET with top gate control. Extracted (c) $g_m$, (d) $\mu_{eff}$, and (e) $R_{SD}$ as a function of $V_{TG}$. Good top gate control is observed. Low on-current is due to SB at S/D as needed for ultra-low power.

Fig. 6 (a) $I_D$ tunability using $V_{BG}$ at fixed $V_{TG}$. (b) Device performance with $L_G$ scaling. (c) $I_{ON}$ versus $1/L_G$ shows a linear increase.

## Real Time Neuron Demonstration and Spiking Neural Network Performance

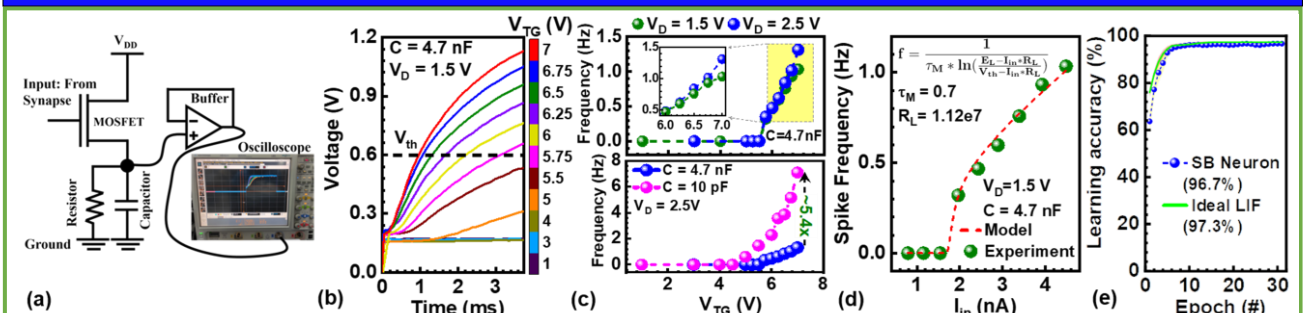

Fig. 7 (a) The circuit used for the neuron demonstration. (b) Measured voltage at capacitor versus time for different $V_{TG}$ at fixed $V_D$ = 1.5 V. (c) Frequency versus $V_{TG}$ of a typical neuron for fixed capacitor (C), different $V_D$, and different C, fixed $V_D$. Frequency tunability is shown with change in $V_D$ and C values. The observed low frequency is real-time to biology enabled by SB ultra-low current. (d) LIF model response fits with the experimental data. (e) Learning accuracy (maximum=96.7%) on Fisher's Iris data set with the proposed neuron model in MATLAB [3].